\documentclass[conference]{IEEEtran}
\IEEEoverridecommandlockouts
\usepackage{cite}
\usepackage{amsmath,amssymb,amsfonts}
\usepackage{graphicx}
\usepackage{textcomp}
\usepackage{xcolor}
\def\BibTeX{{\rm B\kern-.05em{\sc i\kern-.025em b}\kern-.08em
    T\kern-.1667em\lower.7ex\hbox{E}\kern-.125emX}}

\usepackage{algorithm}
\usepackage{algpseudocode}

\usepackage{enumitem}

\usepackage{tikz}
\usepackage{textcomp}
\newcommand\copyrighttext{%
  \footnotesize \textcopyright 2026 IEEE. Personal use of this material is permitted.  Permission from IEEE must be obtained for all other uses, in any current or future media, including reprinting/republishing this material for advertising or promotional purposes, creating new collective works, for resale or redistribution to servers or lists, or reuse of any copyrighted component of this work in other works. 

  Accepted for publication in ICSA-C, International Conference on Software Architecture Companion Proceedings.}
\newcommand{\copyrightnotice}{%
\begin{tikzpicture}[remember picture,overlay]
\node[anchor=south,yshift=10pt] at (current page.south) {\fbox{\parbox{\dimexpr\textwidth-\fboxsep-\fboxrule\relax}{\copyrighttext}}};
\end{tikzpicture}%
}

\usepackage{hyperref}
\hypersetup{
    colorlinks=true,
    linkcolor=blue,
    filecolor=blue,      
    urlcolor=black,
    citecolor=black,
    pdftitle={Overleaf Example},
    pdfpagemode=FullScreen,
    }

\begin{document}

\title{Toward a Sustainable Software Architecture Community: Evaluating ICSA’s Environmental Impact
}

\author{\IEEEauthorblockN{{Mahyar Tourchi Moghaddam}\\
mtmo@mmmi.sdu.dk\\
\IEEEauthorblockA{\textit{SDU Software Engineering} \\
Odense, Denmark}
\and
\IEEEauthorblockN{Mina Alipour}
mial@mmmi.sdu.dk\\
\IEEEauthorblockA{\textit{SDU Software Engineering} \\
Odense, Denmark}
\and
\IEEEauthorblockN{Torben Worm}
tow@mmmi.sdu.dk\\
\IEEEauthorblockA{\textit{SDU Software Engineering}}
Odense, Denmark}
\and
\IEEEauthorblockN{Mikkel Baun Kjærgaard}
mbkj@mmmi.sdu.dk\\
\IEEEauthorblockA{\textit{SDU Software Engineering}}
Odense, Denmark}

%\author{\IEEEauthorblockN
%{The authors are removed to respect anonymity.}}
%{1\textsuperscript{st} Mina Alipour}
%\IEEEauthorblockA{\textit{MMMI Software Engineering} \\
%\textit{University of Southern Denmark}\\
%mial@mmmi.sdu.dk}
%\and
%{2\textsuperscript{nd} Mahyar T. Moghaddam}\\
%\IEEEauthorblockA{\textit{MMMI Software Engineering} \\
%\textit{University of Southern Denmark}\\
%Odense, Denmark \\
%mtmo@mmmi.sdu.dk}}

\maketitle

\copyrightnotice

\begin{abstract}
Generative AI (GenAI) tools are increasingly integrated into software architecture research, yet the environmental impact of their computational usage remains largely undocumented. This study presents the first systematic audit of the carbon footprint of both the digital footprint from GenAI usage in research papers, and the traditional footprint from conference activities within the context of the IEEE International Conference on Software Architecture (ICSA).
We report two separate carbon inventories relevant to the software architecture research community: {\em i)} an exploratory estimate of the footprint of GenAI inference usage associated with accepted papers within a research-artifact boundary, and {\em ii)} the conference attendance and operations footprint of ICSA 2025 (travel, accommodation, catering, venue energy, and materials) within the conference time boundary. These two inventories, with different system boundaries and completeness, support transparency and community reflection. We discuss implications for sustainable software architecture, including recommendations for transparency, greener conference planning, and improved energy efficiency in GenAI operations. Our work supports a more climate-conscious research culture within the ICSA community and beyond.
\end{abstract}


\section{Introduction}

Generative AI (GenAI) has quickly spread across communities, including software architecture. Large Language Models (LLMs) and other GenAI tools are increasingly used for tasks such as code generation, design assistance, and requirements analysis \cite{esposito2025generative}. Recognizing this trend, ICSA 2025 encouraged submissions on {\em Architecture \& Generative AI}, including topics such as the use of LLMs for design decisions, pattern identification, source code generation, design review, and trade-offs\footnote{\url{https://conf.researchr.org/track/icsa-2025/icsa-2025-papers?\#Call-for-Papers}}.
While GenAI claims to improve productivity and innovation, it also incurs high computational costs, especially high energy use, which leads to carbon emissions. Researchers have raised concerns about the environmental impact of AI, e.g., estimating that training a single large NLP model can generate hundreds of tons of CO2-equivalent (CO$_2$eq) \cite{strubell2019energy}. %Even using models at scale can require significant power: a single ChatGPT query is estimated to consume 2-3 Wh of energy, resulting in approximately 2-3 g of CO$_2$eq under normal conditions [CITE]. As GenAI usage increases, its share of digital emissions will also grow.
Studies \cite{jegham2025hungry} and companies' disclosures \cite{altman2025gentlesingularity} show that a typical query requires approximately 0.3-0.5 Wh of electricity, with longer or more complex prompts consuming several watt-hours under realistic serving conditions.

In-person conferences also generate substantial carbon emissions, primarily due to travel. Prior studies show that attendee air travel accounts for the majority of conference emissions \cite{mannheim2025examining}, a finding confirmed by life-cycle assessments across events \cite{neugebauer2020sustainable}. While face-to-face participation remains important for community building, more sustainable planning approaches are needed.

As the ICSA community both adopts GenAI and continues to meet physically, assessing the combined digital and traditional footprint becomes essential. Using ICSA 2025 in Odense, Denmark, as a case study, we measure: {\em i)} emissions from GenAI use in accepted papers, and {\em ii)} emissions from travel, accommodation, venue operations, catering, and materials, based on anonymous data\footnote{\url{https://bit.ly/41D5G8k}}. To date, software architecture conferences have not been systematically assessed in this way, limiting awareness of their environmental impact.

This paper aims to address the gap by conducting a carbon footprint assessment for ICSA 2025. We explore the following research questions: 

\noindent{\bf RQ1:} To what extent did ICSA authors use GenAI, and what is the estimated carbon footprint of this usage? 

%\noindent{\bf RQ2:} To what extent did the ICSA program committee use GenAI, and what is the estimated carbon footprint of this usage? 

\noindent{\bf RQ2:} What are the carbon emissions from traditional conference activities at ICSA 2025?

\noindent{\bf RQ3:} What is the %total carbon footprint of ICSA 2025 when comparing different sources, combining all sources, and what is the
per-capita emission? 

By answering these questions, we quantify digital and physical emissions, identify improvement areas, and propose strategies for more sustainable conferences and research.

%By answering these questions, we aim to quantify digital and physical emissions, identify areas for improvement, and develop strategies to make future conferences and research practices more sustainable.

%%%%%%%
\begin{figure}[ht]
    \centering
    \includegraphics[width=1\columnwidth]{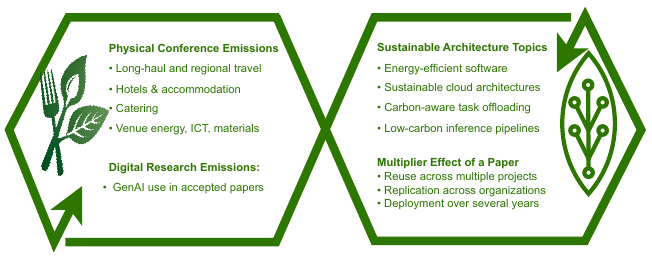}
    \caption{Net-zero framing for the ICSA community.}%: dominant physical and digital carbon sources vs. sustainability-oriented software architecture research as a positive climate lever. How many sustainability-driven software architecture papers are required to compensate for annual ICSA emissions through their real-world deployment impact? This connects the conference footprint directly to a measurable net-zero research objective.}
    \label{met}
\end{figure}
%%%%%%

The contributions are: {\em i)} We present a method for detecting and measuring GenAI tool usage in scientific research outputs. This combines automated detectors with manual verification and energy estimation techniques. {\em ii)} We provide the first empirical estimates of the carbon footprint linked to GenAI in software architecture research. We also include a detailed account of all major conference-related emissions. {\em iii)} We discuss practical recommendations for the software architecture community, including encouraging transparency in reporting computational use, adopting greener research practices, and rethinking conference organization (such as promoting green travel and catering) to reduce carbon impact.

%Our work also adds to a growing conversation on sustainable software architecture. It helps align the community’s practices with broader climate responsibility goals.

%Beyond auditing the emissions and energy consumption of ICSA 2025, this study offers a promising outlook in which software architecture research could be a real actor in addressing climate change. As shown in Figure 1, rather than viewing ICSA’s footprint as a downside, we could see it as a challenge to make a positive impact by encouraging more research on sustainable and sustainability-oriented architectures. For instance, ICSA 2025 published promising research in sustainable architectures, carbon-aware task offloading, and low-carbon inference pipelines. Doing so, the positive effect could outweigh the conference's emissions, towards a net-zero mindset for our community. {\em How many contributions focused on sustainability would be required to compensate for ICSA's annual carbon footprint?} By incorporating this question into our medium- and long-run research planning, we will assume responsibility for the climate through technical innovation, highlighting the role of software architects as active agents in global decarbonization.

Beyond quantifying ICSA 2025’s emissions, this study frames the conference footprint as a measurable baseline for improvement. As Figure 1 shows, rather than viewing it solely as a drawback, it can motivate greater emphasis on sustainability-oriented software architecture research, such as sustainable architectures, carbon-aware task offloading, and low-carbon inference pipelines. An open question is how much sustainability-driven research would be required to offset the conference’s annual carbon footprint. Embedding this perspective into long-term research planning highlights the potential role of software architecture in supporting broader decarbonization efforts. %Here is a link to our %\href{https://docs.google.com/spreadsheets/d/e/2PACX-1vRwZnQbm0D5jdom9yvdWfnX-ydV2KtOf034TAw29DuWnp3jyzBnLsHMbq-njK8EcQ/pubhtml}{\underline {data}}, 
% data included for review purposes: \url{https://bit.ly/41D5G8k}.

The remainder of this paper is organized as follows. Section II presents background on GenAI and conference sustainability. Section III describes the methodology. Section IV reports the results. Section V discusses the findings, Section VI outlines limitations, and Section VII concludes.

%The paper is organized as follows. Section II provides background information by reviewing the rise of GenAI in software engineering, its energy use, and how various activities affect conference sustainability. Section III explains the methods we used for data collection, detecting GenAI usage, and calculating energy and emissions. Section IV shows the quantified emissions and answers research questions RQ1 to RQ4. Section V discusses the findings, and Section VI covers the study's limitations. Section VII acknowledges our AI use, and Section VIII wraps up the paper by emphasizing key findings and future research.

%****Woesrt case scenarios comparison in results***

%%
\section{Background}
%This section provides background on three topics required for a suitable study design and analysis: {\em a)} the rise of Generative AI in software architecture research, {\em b)} the carbon cost of large-scale AI inference, and {\em c)} standard methods for assessing the environmental impact of academic conferences, which together establish the conceptual and methodological foundation for our empirical assessment of ICSA 2025.

%
\subsection{GenAI in Software Architecture: its Good and Bad}

Software architects are increasingly exploring the use of GenAI to support complex design and engineering tasks \cite{nguyen2025generative}. GenAI models, particularly LLMs such as GPT-3/4 and BERT variants, can generate source code, simulate design decisions, and translate requirements into architectures \cite{fan2023large}. For instance, researchers examined whether LLMs have architectural knowledge \cite{dhar2024can}, e.g., the ability to answer design questions or recognize patterns, and assessed how effectively they can generate or refactor architectural artifacts. This is evident in the ICSA 2025 papers discussing the use of GPT for architectural knowledge and component generation \cite{ soliman2025large, shamsujjoha2025swiss, liu2025improving}. The interest in GenAI is due to its ability to manage large amounts of information and automate creative tasks, increasing productivity and supporting decision-making. Several accepted papers at ICSA 2025 highlighted applications such as using LLMs for architecture traceability and generating architectural components with serverless platforms \cite{fuchss2025enabling, arun2025llms}.

Recent ICSA 2025 papers illustrate how GenAI can support more sustainable software architecture. For example, self-adaptive edge-AI approaches aim to improve energy efficiency through hyperparameter optimization and dynamic model adaptation \cite{matathammal2025edgemlbalancer}. Similarly, agent-based frameworks have been proposed to manage energy and resources in data centers by monitoring demand and adjusting cooling or power usage \cite{sera2025dynamic}. These works demonstrate how GenAI-enabled techniques can contribute to reducing software system emissions.

At the same time, awareness of GenAI’s hidden environmental costs is increasing. LLM-based services require substantial computational resources, yet their energy use often remains invisible to end users \cite{barnett2025visualizing}. As researchers increasingly rely on tools such as ChatGPT for writing or analysis, cumulative energy consumption may become significant. Prior discussions have warned that the broader environmental impact of AI can be underestimated amid rapid adoption \cite{inie2025co2stly}. While GenAI offers clear benefits, its energy implications warrant careful consideration.

\subsection{Carbon Cost of GenAI}
%This includes costs and measuring for GenAI

GenAI’s carbon footprint arises primarily from {\em training} and {\em inference}. Training large models is energy-intensive and can generate substantial CO$_2$eq emissions \cite{strubell2019energy}. For example, training GPT-3 required approximately 1,287 MWh of electricity and emitted an estimated 552 tons of CO$_2$eq \cite{patterson2021carbon}. In software architecture research, however, models are typically accessed via APIs or pre-trained checkpoints, making inference the relevant stage for footprint estimation.

Although inference consumes less energy per request, its cumulative impact can be significant. Median energy use per LLM query is around 0.34 Wh, rising to several Wh for long prompts, and large-scale or repeated use can accumulate rapidly. As model sizes and adoption grow, inference-related energy consumption may rival or exceed training \cite{desislavov2023trends}.

Accurately measuring inference emissions is challenging because proprietary cloud models obscure hardware and energy details. Researchers therefore rely on proxy measurements or emission calculators. For example, prior work replicated typical GenAI tasks on local GPUs and measured energy with Carbontracker \cite{inie2025co2stly}, finding that fine-tuning tasks were more carbon-intensive than prompt-based usage. Limited transparency and inconsistent reporting further complicate precise estimation.

\subsection{Measuring Sustainability of Events}

Sustainability research increasingly examines the carbon footprint of academic events. Life-cycle assessments comparing in-person, hybrid, and online formats consistently show that in-person conferences have the highest emissions, largely driven by long-distance air travel \cite{tao2021trend, cavallin2023life}. Other sources, such as venue energy, accommodation, materials, and local transport, contribute smaller shares \cite{van2021emission}.

The Hotel Carbon Measurement Initiative (HCMI) provides standardized estimates for hotel emissions, typically 5--15 kg CO$_2$eq per room-night for operational impacts \cite{HCMI2020}, potentially rising to around 20 kg CO$_2$eq when indirect effects are included \cite{filimonau2011reviewing}. Venue emissions depend on building characteristics and the local energy mix. Assessing ICSA 2025 therefore establishes a baseline to identify high-impact mitigation opportunities.

%
%\subsection{}

%%%%%%%
\begin{figure}[ht]
    \centering
    \includegraphics[width=.8\columnwidth]{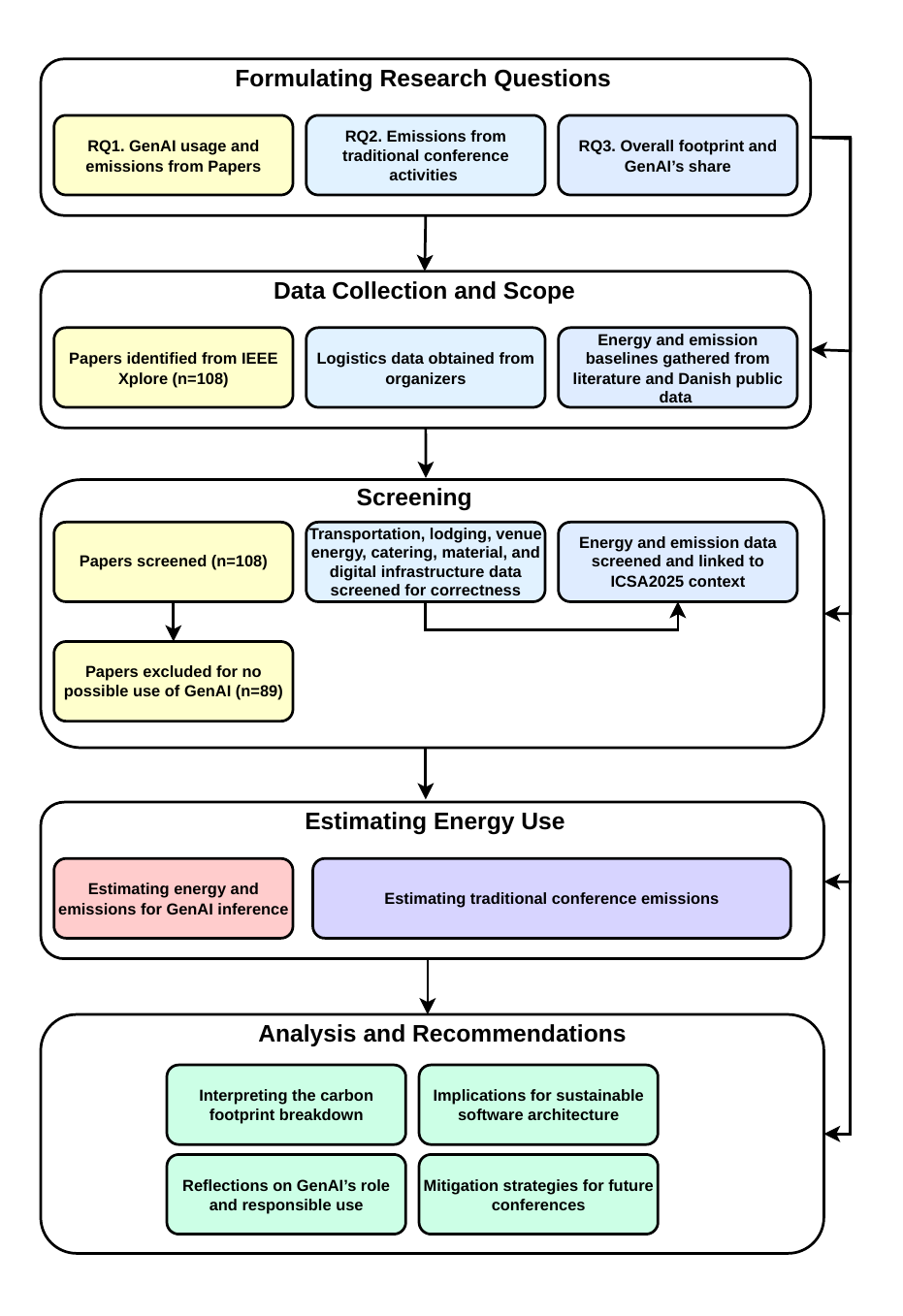}
    \caption{Research methodology.}
    \label{met}
\end{figure}
%%%%%%

%%
\section{Research Methodology}

%Research questions (that are addressed with the results. E.g.:
%1. How was the use of GenAI in ICSA 2025?
%2. What was the CO2 generation of accepted papers at ICSA 2025?
%3. How much CO2 did other conference activities generate?
%4. The commutative CO2 generated from the event
%Then we provide recommendations.

%
%\subsection{Screening and Selection Strategy}

%Requires assurance with the steps followed (including identification, screening, and inclusion).

%We may also need inclusion and exclusion criteria.

%We developed a methodology to address our research questions by identifying GenAI usage in ICSA 2025 papers, estimating associated energy consumption and CO$_2$eq emissions, and assessing emissions from key conference activities. Our approach prioritizes transparency and replicability, with all assumptions documented for future updates.

%%
\subsection{Data Collection and Scope}

{\bf Papers Data:} We analyzed the 108 accepted ICSA 2025 papers (28 main track, 80 companion) to ensure a manageable and replicable scope. Ethics and data protection are discussed in Section III-F.

{\bf Activity Data:} To estimate emissions from conference logistics, we collected organizer-provided data on 229 in-person attendees, their geographical distribution and registration types, the conference schedule, venue details at the University of Southern Denmark (room sizes and usage), and the catering menu.

We considered the following emission sources:

\begin{itemize}[leftmargin=*]
\item {\bf \em GenAI in Papers:} Inference-related computing for GenAI (text/image generation). Training and traditional computation were excluded.
\item {\bf \em Transportation:} Travel to and from Odense (primarily air, train, or car), including minor local transfers.
\item {\bf \em Accommodation:} Hotel stays based on registration category, using HCMI-aligned per-night factors.
\item {\bf \em Venue Energy:} Electricity for lighting, HVAC, audiovisual systems, and digital infrastructure, estimated from venue size and typical energy intensity.
\item {\bf \em Catering:} Meals and refreshments calculated using life-cycle carbon intensity factors.
\item {\bf \em Materials and Waste:} Printed materials, conference items, and waste, estimated using standard LCA factors.
\end{itemize}

All emissions are reported in kg CO$_2$eq using region-specific emission factors (e.g., Denmark’s electricity grid intensity).

\subsection{Detecting GenAI Usage}

Identifying how authors use GenAI in research is challenging, as they often do not clearly state its use. We conducted our study using automated text analysis and a manual review as follows:

\noindent {\bf \em 1. Keyword Search:} We scanned the papers for specific terms related to GenAI, such as “ChatGPT”, “GPT-3”, “GPT-4”, “large language model”, “Transformer”, “BERT”, “Generative AI”, “LLM”, “OpenAI API”, “prompt”, “fine-tune” to highlight relevant sections. Mentions of papers and tools linked to GenAI also indicated potential usage.

\noindent {\bf \em 2.  AI-Generated Text Detection:} We obtained access to the GPTZero\footnote{\url{https://gptzero.me}.} API for research use to determine whether parts of the papers might be AI-generated \cite{adam2026gptzerorobustdetectionllmgenerated}. Because results can be unreliable, we considered a paper as possibly containing AI-written text only if the entire paper was flagged by the detectors with at least 50\% confidence.
%GPTZero uses a deep-learning model to analyze text features like “perplexity” and “burstiness” to identify human-written v.s. AI-generated content. It reports achieving approximately 98.6\% accuracy in detecting AI text and a 0\% false positive rate for human content. For mixed documents, accuracy is reported to be 96.5\% with under 1\% false positives. As a probabilistic tool, its reliability increases with longer texts and serves as an indicator of authorship rather than definitive proof \cite{gptzero2024benchmark}.

\noindent {\bf \em 3. Manual Content Analysis:} We reviewed the methodology sections to identify applications of GenAI, using a coding scheme to track possible research stages and purposes. We looked for quantitative data to estimate energy.

After completing these steps, we created a list of papers that likely used GenAI. If a paper showed no signs of GenAI or was detected with less than 50\% probability, we assumed zero GenAI-related emissions. If it mentioned GenAI without details, we estimated possible usage scenarios.

\subsection{Estimating Energy and Emissions for GenAI Inference}

In our assessment of the carbon footprint from generative AI usage at ICSA 2025, we used EcoLogits\footnote{\url{https://ecologits.ai/latest/}} \cite{rince2025ecologits}, a tool that estimates the energy use and greenhouse gas emissions from LLM inferences. EcoLogits tracks energy consumption and environmental impacts from generative AI models accessed via API calls, providing metrics like energy consumed and carbon emissions.
The impact of one request is:

\[
I_{\text{request}}
= E_{\text{request}} F_{\mathrm{em}}
+ \frac{\Delta T}{\Delta L} \, I_{\text{servere}}^e ,
\]

where $E_{\text{request}}$ is the energy used by the server (including an overhead for cooling) to serve the request, $F_{em}$ is the emission factor of the electricity (kgCO$_2$eq per kWh) at the data center, $\Delta T/\Delta L$ is the fraction of the server’s lifetime consumed by this inference, and $I_{server}^e$ is the total embodied GHG impact of the server hardware.

For our analysis, we applied EcoLogits to each instance of generative AI use in the ICSA 2025 submissions, e.g., using ChatGPT for tasks like text generation. Key inputs included the model used, the output token count, server hardware, data center location, and efficiency factors. 
By inputting these parameters, EcoLogits helped us calculate the energy consumption and carbon emissions for each usage event. We then totaled the emissions across all events to quantify the conference-wide impact.
It is important to note that this estimate is based on certain assumptions, particularly for models with limited publicly available information. The details of deployment can vary: factors such as hardware efficiency and the data center's carbon intensity can influence the results. We used standardized inputs to minimize inaccuracies.

\subsection{Traditional Conference Emissions}
\subsubsection{Transportation}
When we look at the environmental impact of in-person conferences, the largest contributor is travel. To estimate how much CO$_2$eq each participant generates from their travel, we first assessed their emission by their origin country and further categorized their journeys as local, neighbor, regional (within Europe), or long-haul (intercontinental). For local and neighborhood travelers, we assumed they primarily use trains, while those coming from long distances would typically opt for flights.
When it comes to flights, we used specific emission factors: approximately 0.15 kg of CO$_2$eq per passenger-kilometer for shorter flights and 0.11 kg/km for longer flights \cite{defra2023aviation}. The impact of non-CO$_2$eq emissions occurring at high altitudes should also be included. The exact calculation was performed using {\em myclimate} calculator that considers all possible contributing variables \cite{myclimate2025flight}.
Train travel was assigned a much lower emissions factor of 0.024 kg/km per passenger \cite{carbonlabel2025data} specified for Denmark, considering that most of Europe's train network is powered by electricity. On the other hand, car travel for a few local attendees was estimated at 0.137 kg/km, which is typical for a single-occupancy petrol car \cite{carbonlabel2025data}. The overall transportation impact was simply calculated by adding up the emissions from all individual trips made by attendees:

\begin{equation}
C_{\text{transport}} \;=\; \sum_{i \in P} d_i \,\times\, EF_{\text{mode}(i)}~,
\end{equation}

where $d_i$ is the round-trip distance for participant $i$ and $EF_{\text{mode}(i)}$ is the emission factor for their mode of travel. This accounted for a dominant share of the conference’s carbon footprint, as reported in Section IV.

\subsubsection{Accommodation}
For each non-local attendee, we estimated emissions from hotel stays during the conference. We assumed an average of 4 nights per main conference attendee, 6 nights for full conference attendees, and 3 nights for workshop attendees. We adopted the HCMI factor of approximately 5.87 kg CO$_2$eq per room-night as a representative average for hotel stay emissions in Denmark \cite{climatiq2025ef}. Therefore, if $N_{\text{hotel}}$ is the total number of attendee-nights in hotels, the accommodation footprint is:  
\begin{equation}
C_{\text{accommodation}} \;=\; N_{\text{hotel}} \times 5.87~\text{kg CO$_2$eq/night}~
\end{equation}

This yields on the order of tens of kilograms CO$_2$ per attendee based on their registration category. Accommodation appears as a significant but smaller source than travel.

\subsubsection{Venue Energy}
The conference took place in university buildings, thereby consuming energy for electricity, heating, and ventilation. To estimate the venue’s energy consumption, we considered the venue's size (approximately 1000 square meters, including auditoriums and classrooms) and the event duration (8 hours per day for 5 days). Using typical energy intensity values for modern academic buildings \cite{merabtine2018building}, we calculated the total electricity required for lighting, AV equipment, Wi-Fi, and heating or cooling. Our estimates, together with the venue's technical service, suggested that the overall energy demand (including digital infrastructure) for the conference was around 2,300 kWh.
Next, we accounted for the carbon intensity of the local grid, which in Denmark is approximately 0.173 kg of CO$_2$eq per kWh \cite{nowtricity2025denmark}. This allowed us to estimate the venue's carbon emissions at 397.9 kg CO$_2$eq. In formula form, the venue footprint could be expressed as $C_{\text{venue}} = E_{\text{venue}} \times EF_{\text{grid}}$. %, but in sum, this category contributed only around $1\text{-}2\%$ of the total emissions. {\bf EDIT TEXT based on results}
Considering the registrations and their categories, a person's share of venue-related energy use is 3.03 kWh/day, and venue-related emissions are 0.523 kg CO$_2$eq/day.

\subsubsection{Catering}
When assessing the environmental impact of conference catering, we can break it down by considering the CO$_2$eq emissions per person per meal. For instance, a full conference attendee would have approximately five lunches, one reception, and one banquet dinner over the course of the event, along with coffee breaks each day.
For lunch (30\% vegetarian and 70\% meat-based), we calculated roughly 2.5 kg of CO$_2$eq per person \cite{fvm2021climateisserved, denStoreKlimadatabase2025}, while the richer banquet dinner was estimated at 5 kg CO$_2$eq, and reception 3.5 kg CO$_2$eq. ICSA 2025 also served a light dinner (sandwiches) on 3 evenings, each of which accounted for 0.85 kg CO$_2$eq. The lighter refreshment breaks (coffee and cake/snacks) contribute about 0.3 kg CO$_2$eq each. When we add these together, one attendee's total food-related emissions for the event are approximately 26.55 kg of CO$_2$eq if they attended the full conference.
For full conference participants, if we multiply this by the number of attendees (\(N\)) in that category, we can use the following formula to estimate the catering emissions for a full conference attendee (other categories will have reduced days): \(C_{\text{catering}} = N \times (5 \times 2.5 + 1 \times 5 + 1 \times 3.5 + 10 \times 0.3 + 3 \times 0.85)\).
%\[
%C_{\text{catering}} = N \times (5 \times 2.5 + 1 \times 5 +  1 \times 3.5 + 10 \times 0.3 + 3 \times 0.85)\
%\]

%\text{kg CO$_2$e}~,

This formula will be applied to all other registration categories, in which the number of days and the number of attended events vary. %This can lead to a total of a few thousand kilograms of CO2e.  %{\bf EDIT} With our 229 attendees, considering all the possible cases, we would see around {\bf XXXX Final Calculation} tons.
We acknowledge that actual emissions may differ slightly due to special dietary options provided, but catering typically accounts for a smaller share of the event's total emissions.

\subsubsection{Materials and Waste}
Emissions from conference materials and waste were relatively lower. We included the footprint of printed materials (badges, posters) and any conference swag, as well as waste disposal. Based on rough estimates made together with the organizers' technical service, %we allocated a few kilograms of CO$_2$e per attendee for all materials and waste. For instance, 
printing and paper use per person emitted approximately 0.3 kg of CO$_2$, attendee swag another 2.3 kg, and waste (food packaging, etc., not recycled) ~0.5 kg. In aggregate, we estimate {\em Materials \& Waste} contributed $\sim$709.9 kg CO$_2$eq.

\subsection{Assumptions}
Our objective is to provide a transparent, reproducible baseline of ICSA’s footprint and identify the dominant emission drivers at an order-of-magnitude level. As is common in conference footprinting \cite{tao2021trend, epa2018greenhouse}, some activity data (e.g., travel mode choice) are not observable post hoc without dedicated surveys, and therefore must be modeled using explicit assumptions.
To avoid overstating precision, we mentioned all key assumptions and considered scenario bounds for parameters that materially affect results, such as the share of rail vs air travel in neighbor and regional categories. %We additionally conduct a sensitivity analysis to verify that the headline qualitative result—travel as the dominant source—holds across all plausible scenarios.

\subsection{Ethics, Governance, and Data Protection}

This study is a meta-research assessment of the environmental footprint of a scientific event and related research artifacts. It does not involve interventions, experiments, or interaction with human participants. To minimize privacy risks, we analyzed only: {\em i)} publicly available bibliographic/paper content, and {\em ii)} aggregated, anonymized conference statistics. We did not receive, store, or process direct identifiers (such as names, addresses, emails, or any identities).
For conference attendance and logistics, we followed GDPR-aligned data minimization and purpose limitation principles and restricted our inputs to the least granular data necessary for carbon accounting \footnote{\url{https://eur-lex.europa.eu/eli/reg/2016/679/oj/eng}}. Where any underlying administrative data were used to produce aggregate statistics, these were handled under SDU’s institutional governance processes for research and data protection (including the use of SDU-approved storage and access control, and deletion or anonymization after project completion) \footnote{\url{https://sdunet.dk/en/research/legal-services/researchportal_legal_gdpr}} previously obtained for ICSA 2025.
In line with IEEE publication policies for work involving human-related data \footnote{\url{https://conferences.ieeeauthorcenter.ieee.org/author-ethics/guidelines-and-policies/submission-policies/}}, we include this explicit ethics and data protection statement and clarify that no individual-level personal data were processed in our analysis and reporting.

\section{Results}
%We now present the results of our carbon footprint analysis, organized by the research questions.

\subsection{GenAI Usage in Papers (RQ1)}

Out of the 108 papers at ICSA 2025, we identified 26 papers (24\%) that used Generative AI.
This moderate but notable share reflects its emerging but not yet dominant role in software architecture research.
Only one paper reported their use of GenAI as an assistant in content generation. Our methodology also flagged 11 additional papers with a probability of GenAI use below 50\%, which were excluded from emission auditing. It is possible that some authors used GenAI indirectly, e.g., to get feedback and ideas, which our analysis cannot confirm with certainty.

In addition to text generation, the identified use cases of GenAI in ICSA papers include:
{\em i)} architecture design assistance, where models like GPT-4 and GPT-3 are used to generate and evaluate design-related code stubs through interactive queries; {\em ii)} requirements traceability and analysis, e.g., a paper leveraging a pre-trained LLM to extract architecture component names from text, utilizing named-entity recognition, which required considerable computation; {\em iii)} AI as subject, e.g., a paper studied a GenAI model, likely GPT-3.5 or GPT-4, by asking it software architecture questions, focusing solely on inference-based responses without modifying the model; {\em iv)} code generation and testing, where, e.g., a study indicated the use of a generative code assistant for automating the creation of alternative code implementations to test energy efficiency. These papers demonstrate how LLMs enhance design, analysis, research, and code generation in software architecture, all of which are necessary to advance the field.

Across the GenAI-using papers, no instance of training a large model from scratch was found. All usages were via API, fine-tuning smaller models, or using open pre-trained checkpoints. This suggests that most software architecture researchers focus primarily on using generative AI for inference rather than on creating or training generative AI models. As a result, the environmental impact, in terms of carbon emissions, is largely limited to the inference process rather than the entire development phase.
The explicit mention of GenAI use was detected in the papers' evaluation phase (e.g., comparing LLMs with human performance), the implementation phase (e.g., generating artifacts), and the data analysis phase (e.g., traceability), and as previously mentioned, only one paper explicitly mentioned using GenAI for literature review or writing.
%
%
%
%Overall, GenAI played a role in about one-quarter of the research works, indicating an emerging presence but not yet ubiquity in software architecture research.
%
Overall, GenAI played a role in more than one-quarter of the research works, indicating an emerging presence but not yet ubiquity in software architecture research.

In terms of emissions, the CO$_2$eq of papers ranged from 176 g CO$_2$eq (with 363 Wh of energy consumption) to 524 g CO$_2$eq (with 1077 Wh of energy consumption). 
The total emissions from papers' GenAI use were 8694 g CO$_2$eq (with 17827 Wh of energy consumption), which would have been 36720 g CO$_2$eq (with 75168 Wh of energy consumption) if all papers used GenAI to generate their content.
The results confirm that, while increasing awareness is crucial, GenAI-generated emissions from research are far lower than those of traditional conference activities.

\subsection{Emissions from Traditional Conference Activities (RQ2)}

The carbon emissions from the traditional physical components of ICSA 2025 far exceeded those from GenAI. {\bf \em Travel} was by far the dominant contributor, accounting for approximately 94\% of total emissions, 172.218 tonnes of CO$_2$eq just from attendee transportation. This aligns with prior studies that found that attendee air travel accounts for most of conference emissions. However, the share of emissions from travel is higher for ICSA 2025 due to the high number of intercontinental attendees, and as Denmark is relatively sustainable in other conference activities. The most emitting activity after travel was lodging, with 5.89 tonnes of CO$_2$eq, approximately 3\% of total emissions.
{\bf \em Catering} (meals and beverages) was on the order of 2\% of emissions with 4.26 tonnes of CO$_2$eq, {\bf \em Materials and waste} emitted 710 kg CO$_2$eq and  {\bf \em Venue} emitted 398 kg CO$_2$eq.

%\begin{figure}[ht]
%    \centering
   % \includegraphics[width=.8\columnwidth]{Travel.jpg}
   % \caption{Considerable share of "travel" in ICSA 2025 total estimated emissions.}
   % \label{met}
%\end{figure}

%For RQ2, we find that the physical aspects of the conference (especially travel) generate a considerable carbon footprint.

\begin{figure}[ht]
    \centering
    \includegraphics[width=1\columnwidth]{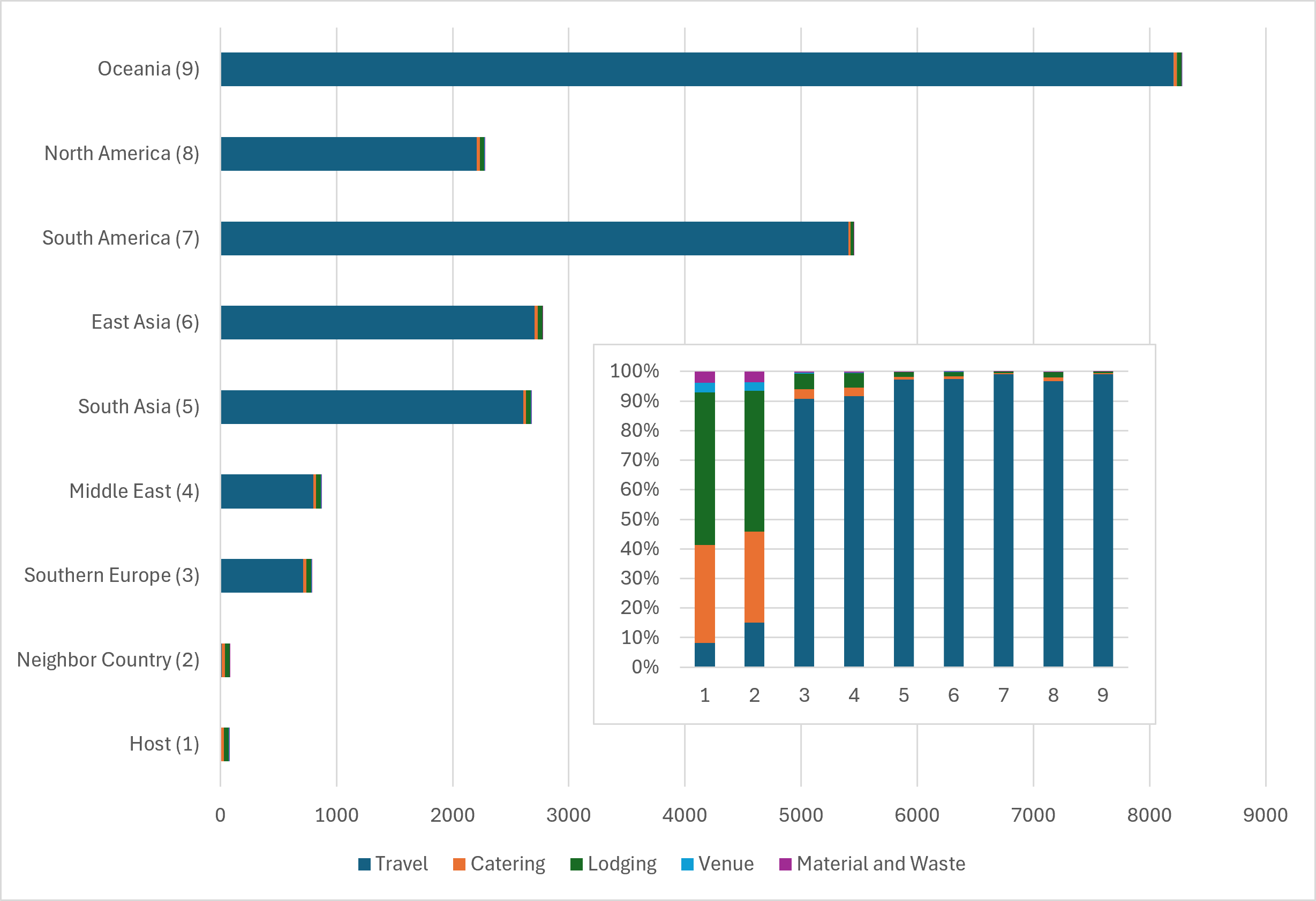}
    \caption{The comparison of worldwide regions participants (full conference) emissions (kg CO$_2$eq) and the share of various emission contributors for each region. The smaller chart shows the percentage of emissions by region, indicating that, for local and neighboring countries, participants' lodging and catering dominate, whereas for others, travel accounts for the largest share.}
    \label{met}
\end{figure}

\subsection{Overall Footprint (RQ3)}

By aggregating all available data from the calculations above, the estimated total carbon footprint for ICSA 2025 is around 183 t CO$_2$eq. With 229 in-person attendees, this corresponds to approximately 0.8 t CO$_2$eq per participant, comparable to the emissions from a round-trip air travel from southern Italy to Denmark. This per-person share is quite high for a brief event, largely attributable to a significant proportion of long-distance travel. %Crucially, our analysis finds that the vast majority of emissions originated from traditional physical activities, with travel alone accounting for approximately 80\%. By contrast, the Generative AI (digital) component was less than 1\% of total emissions: we roughly estimate XXX hundred kilograms of CO2e linked to all GenAI usage by authors and reviewers.
In Figure 3, the larger bars show CO$_2$eq associated with participants traveling from local, neighboring, and intercontinental regions, and the inset specifies the composition of each region’s contribution. Long-haul travel considerably increases the emissions of intercontinental attendees compared to those from Europe or Denmark. For participants located closer to Odense, travel emissions are relatively minor, and hotel stays and catering become the primary sources of carbon emissions for both local and neighboring delegates.

%{\em RQ4 can thus be answered clearly:} the overall carbon footprint of ICSA 2025 was overwhelmingly dominated by conventional sources, with GenAI-related emissions playing a minimal role. This confirms that, at present, facilitating face-to-face collaboration incurs a substantially higher carbon cost than the emerging digital tools used in that collaboration.

%While GenAI’s footprint at ICSA 2025 is currently insignificant, we emphasize that this finding is contingent on context. It highlights where the biggest leverage lies for emission reductions (in travel and other logistics), rather than suggesting that GenAI’s impact be ignored. %If the use of GenAI in research were to increase dramatically, the calculus could change. We can envision a scenario where, say, half the papers involve training or running large models on cloud infrastructure: in such a case the GenAI portion might rise to several tons of CO2. Even then, unless attendance-related emissions drop, travel would likely remain the dominant factor. 
Our results thus suggest a current asymmetry: the environmental cost of disseminating research far exceeds that of conducting it, at least in software architecture, where computational demands are moderate. This may not hold in fields with extreme AI computation (where research itself can emit tens of tons), but in our case, it is a striking insight. Going forward, it will be essential to monitor GenAI’s footprint, as small does not mean zero, yet the immediate priority for emissions reduction lies in rethinking conference logistics.

%Finally, to place ICSA 2025’s footprint in a broader context, we consider the best- and worst-case scenarios for comparison with our actual findings. Under the best-case assumptions (a regionally focused, low-impact event), ICSA’s total emissions might have been as low as around 20 t CO2e, whereas under worst-case conditions (maximal long-haul travel and inefficiencies), it could exceed 300 t CO2e. Our actual 87 t footprint lies between these extremes, closer to the lower end, indicating that some mitigation is already in effect. In all scenarios, the fraction of GenAI remains minor.

\section{Discussion}
%The above results provide a clear picture of the environmental impact of ICSA 2025 and the major sources of its CO$_2$ emissions. In this section, we discuss the implications of these findings for the software architecture community and propose strategies for reducing the carbon footprint of future conferences and research practices.

\subsection{Interpreting the Carbon Footprint Breakdown}
%Our analysis confirms that [CITE] in-person conferences have a carbon footprint heavily dominated by long-distance travel. Approximately four-fifths of ICSA 2025's emissions came from attendee transportation. This 

The high share of transportation emissions highlights that any meaningful reduction in a conference's carbon footprint must involve addressing travel. Improvements in other areas are beneficial with comparatively smaller gains.
The ICSA Steering Committee can thus consider models such as {\em regional hubs} to reduce travel, which might yield far greater carbon savings than any on-site operational tweaks. To bend the emissions curve, decisions about where and how we meet are paramount. %, as evidenced by the dramatic footprint differences between our best- and worst-case projections.
That said, smaller changes should not be ignored; measures such as greener catering or venue-level renewable energy use still align with sustainability values and can be implemented at low cost, thereby strengthening a culture of environmental mindfulness. 
%While no plausible increase in GenAI usage could offset the carbon benefits of reducing travel, 
Furthermore, an awareness of AI's impact can promote greater responsibility as its use continues to grow.

\subsection{Implications for Sustainable Software Architecture}
The software architecture research community can draw several lessons from these findings. First, there is a strong case for greater {\em transparency in reporting computational resources and emissions} in our research. In our study, we had to infer GenAI usage post-hoc, but going forward, authors could be encouraged (or required) to disclose any extensive use of AI models or cloud computing and even estimate the associated energy/carbon footprint. This effort could align with best practices in other fields like machine learning that have introduced paper checklists, including environmental impact disclosures. Making such reporting a norm would raise awareness and encourage researchers to consider efficiency in their methodology. Second, our community might widen its notion of what constitutes good software architecture practice to include sustainability considerations. Just as architects design systems with quality attributes in mind, we should treat {\em carbon footprint as a relevant quality attribute} of both our research processes and the systems we study. Embracing this perspective, ICSA and similar conferences could innovate in how they are structured, or better said, {\em architected}, exploring distributed or hybrid conference models that maintain scholarly exchange while dramatically cutting emissions.
Lastly, the findings reinforce that to improve sustainability, our primary focus should be on the {\em big-ticket items} (e.g., logistics) when planning events. At the same time, as GenAI becomes more common in our research, we should promote efficient use of computing (e.g., use smaller models or local resources when feasible) so that the digital side of research remains as sustainable as possible.

\subsection{Mitigation Strategies for Future Conferences}
Building on our results, we outline a few strategies that could reduce the carbon footprint of future ICSA editions:

\noindent{\bf Hybrid conferences:} as ICSA maintains strong community ties, a physical meet-up is necessary. However, even partial virtualization helps, e.g., alternating in-person and hybrid conference years, or hosting a single event with both physical and virtual tracks, could cut the travel footprint by a large margin. %(potentially 50\% or more).

\noindent{\bf Regional Meetups and Co-located Events:} instead of one global gathering, the conference could adopt a distributed model (multiple regional hubs linked by video) or coordinate timing with other conferences. Presumably, if ICSA is scheduled right after another major conference in the same area, international attendees could get the chance to join both events in one trip. This could cut down on their flight emissions for each event. Also, offering regional workshops or local mirror sites could be a great option, making it easier for participants to attend something closer to home.
 
\noindent{\bf Low-Carbon Travel Incentives:} encourage and facilitate travel by train or other low-carbon means for those who are within a reasonable distance. Organizers can support more sustainable travel by offering incentives like registration discounts for train travelers or by facilitating local attendees to coordinate ride-sharing. Including carbon offsets in the registration fee or through sponsors can also help balance out travel emissions, while it is essential to handle offsets transparently and responsibly.

\noindent{\bf Sustainable Catering and Accommodation:} adopt plant-forward catering menus (vegetarian or vegan by default), since meals with less meat can substantially lower food-related emissions. Similarly, partner with hotels that have sustainability certifications or use renewable energy, and inform attendees about greener lodging options. These steps ensure that the necessary emissions from food and lodging are minimized.

\noindent{\bf Reducing Materials and Waste:} continue phasing out printed materials by using digital programs and conference apps. Avoid disposable swag or replace it with eco-friendly or non-physical tokens (e.g., donating to an environmental cause on behalf of attendees). While these actions have a small direct carbon impact, they are symbolically important and contribute to a culture of sustainability.

\noindent{\bf Greener Research Practices:} promote energy-efficient computing in research. For example, the conference could introduce a recognition (an informal {\em Green Paper Award}) for work that achieves its results with remarkably low resource usage or by using innovative efficiency techniques. Encouraging authors to use renewable-powered infrastructure or to share their compute footprint in submissions can gradually make {\em green AI} a part of the community ethos.

%Implementing a combination of these measures could enable ICSA to reduce its carbon footprint significantly in future years. 
%Importantly, many of these changes can be made without sacrificing the core benefits of the conference. 
%The social value of face-to-face interaction is undeniable, but as we innovate with smarter planning, we can strive to {\em maximize the benefits of gathering while minimizing the environmental costs}. 
%The software architecture community has an opportunity to lead by example, showing how a high-profile conference can adapt to the climate crisis through thoughtful design and policy choices.

\subsection{Reflections on GenAI's Role and Responsible Use}
GenAI's growing prevalence in research requires forward-looking consideration. If, in the near future, a much larger fraction of papers leverage GenAI or conduct compute-intensive experiments, the cumulative emissions could increase. The community should be proactive in establishing norms for {\em responsible AI use}. This could include expectations to disclose usage and to favor energy-efficient modeling techniques when possible. Introducing an environmental lens to research ethics is a novel but important idea; for instance, if a certain evaluation would consume an extraordinary amount of energy for marginal scientific gain, should researchers rethink it? By making energy impact a consideration in research design, we encourage innovation in more sustainable methods. %Similarly, guidelines may be needed for {\em reviewers' use of AI}, e.g., conferences might highlight the rules that reviewers cannot generate evaluative content, emphasizing confidentiality, and ensuring that human experts remain accountable for review quality. 
Finally, it is worth exploring ways that GenAI itself could contribute to sustainability solutions. AI tools might assist in optimizing conference planning (e.g., smarter scheduling to minimize travel) or help architects design more energy-efficient software systems. As we integrate GenAI into our workflows, we should also incorporate sustainability thinking into how we use these tools. The goal is to reap the benefits of GenAI-driven innovation while keeping its energy footprint in check.

\section{Limitations}

While this study aims to raise awareness, several limitations apply:

\begin{enumerate}[leftmargin=*]
    \item Many estimates rely on assumptions and proxy data (e.g., modeled GenAI usage), introducing uncertainty. The total footprint may vary by approximately $\pm$20\%, though this does not affect the main conclusions.
    
    \item Emission factors differ across sources and methodologies (e.g., flights, accommodation, catering, and grid intensity). We used representative averages; actual values may vary depending on context and specific providers.
    
    \item The analysis focuses on major emission sources and excludes some indirect impacts (e.g., hardware manufacturing, full life-cycle effects, and shared travel for multi-purpose trips). Some data center overhead may not be fully captured.
    
    \item Detecting GenAI usage in papers is imperfect; subtle or minor uses may have gone unnoticed. However, substantial usage is unlikely to have been systematically missed.
    
    \item We evaluated environmental impact only and did not assess economic, social, or research-quality implications of proposed mitigation strategies.
    
    \item The study considers carbon footprint (CO$_2$-equivalent) only and does not include other environmental dimensions such as water use or e-waste.
    
    \item As a mid-sized software engineering conference, ICSA 2025 may not represent larger or more compute-intensive events.
    
    \item GenAI inference estimates rely on limited publicly available data; future transparency may enable more precise calculations.
\end{enumerate}

Despite these limitations, the analysis provides a transparent and sufficiently robust baseline to support the study’s conclusions and recommendations.

\section{Acknowledgment of AI Use}
In this paper, we made limited and transparent use of AI-supported tools. Grammarly was used for proofreading, editing, shortening, rephrasing, and refinement. We used ChatGPT with a 38-word prompt to experiment with two papers totaling 21450 words, contributing 4.83 g CO$_2$eq. We utilized the AI detection tool (GPTZero) to examine the ICSA 2025 papers for any AI-generated content. All the analytical reasoning, methodological choices, and significant contributions presented in this paper are solely those of the authors. %We believe that being transparent about AI use aligns with the growing recommendations of research communities.

%\vspace{-.7em}
\section{Conclusion}
We presented the carbon footprint assessment of ICSA 2025, including the traditional conference emissions and the impact of GenAI usage in research. Our analysis found that the conference’s total emissions were approximately 183 t CO$_2$eq, with roughly 94\% attributable to attendee travel. Improving sustainability in the ICSA community depends on reducing travel and greening our event planning, as well as optimizing researchers’ use of AI tools. The carbon cost of all GenAI-related activities at the conference was 8694 g CO$_2$eq. We thus encourage transparent and efficient use of computational resources as GenAI becomes more common, so that its footprint remains minimal. %We hope that this work provides a data-informed basis for the community to reflect on its practices and to take concrete steps towards a more sustainable future for software architecture research.

%%
%\subsection{Traditiona Emissions Calculation}

%%%%
%\subsubsection{Transportation}

%%%%
%\subsubsection{Accommodation}

%%%%
%\subsubsection{Venue Energy Usage}

%%%%
%\subsubsection{Materials and Waste}

%%%%
%\subsubsection{}

%
%\subsection{Assumptions and Estimations}

%
%\subsection{CO2 Generation of GenAI Calculation}
%Methods with a picture, when including paper screening for keywords, using GPTZero, and further using the EcoLogits for estimation of energy.

%
%\subsection{CO2 Generation of other conference activities}
%Including participants Transport, accommodation, Venue energy, food and catering, material and waste emissions, etc. etc.

%
%\subsection{Mthods}

%
%\subsection{Aggregation and Metrics}

%%
%\section{Results}

%
%\subsection{Answer to Q1}

%
%\subsection{Answer to Q2}

%
%\subsection{Answer to Q3}

%
%\subsection{Answer to Q4}

%The cumulative CO2 generation estimation

%%
%\section{Discussion}

%You may have a figure about the most common use of GENAI in SA, having both the amount of use as well as the CO2 generated. Use could be low but generation high, in e.g., model training...

%
%\subsection{Rationals}

%
%\subsection{Responsibilities}

%
%\subsection{How Make ICSA Greener}

%%
%\section{Threats to Validity}

%%
%\section{GenAI Disclosure}

%%
%\section{Conclusion}

%%
%\section{}

%%
%\section{}

%%
%\section{}

%%
%\section{}

%\vspace{-.5em}
\bibliographystyle{IEEEtran}
\bibliography{bib}

\end{document}